\documentclass{tlp}

\usepackage{xspace}
\usepackage{epsfig}
\usepackage{graphicx}
\usepackage{latexsym}

\def\ii#1{\hbox{\it #1\/}}
\def\is#1{\hbox{\scriptsize\it #1\/}}
\def\rs#1{\hbox{\scriptsize #1}}

\def\beq{\begin{equation}}
\def\eeq#1{\label{#1}\end{equation}}
\def\ba{\begin{array}}
\def\ea{\end{array}}

\def\bdm{\begin{displaymath}}
\def\edm{\end{displaymath}}
\def\bea{\begin{eqnarray}}
\def\eea{\end{eqnarray}}
\def\up{\!\!\uparrow\!\!}
\def\upp{\!\!\uparrow}

\newtheorem{prop}{Proposition}

\begin{document}

\title[Temporal Phylogenetic Networks and Logic Programming]
{Temporal Phylogenetic Networks\\ 
and Logic Programming}

\author[Esra Erdem, Vladimir Lifschitz, and Don  Ringe]
{ESRA ERDEM \\
Institute of Information Systems 184/3\\
Vienna University of Technology, A-1040 Vienna, Austria \\
\email{esra@kr.tuwien.ac.at} 
\and
VLADIMIR LIFSCHITZ \\
Department of Computer Sciences   \\
University of Texas at Austin, Austin, TX 78712, USA \\
\email{vl@cs.utexas.edu}
\and
DON RINGE \\
Department of Linguistics \\
University of Pennsylvania, Philadelphia, PA 19104, USA \\
\email{dringe@unagi.cis.upenn.edu} }

\date{}

  \submitted{30 November 2004}
  \revised{30 May 2005}
  \accepted{25 August 2005}

\maketitle

\begin{abstract}
The concept of a temporal phylogenetic network is a mathematical model of
evolution of a family of natural languages.  It takes into account the
fact that languages can trade their characteristics with each other when
linguistic communities are in contact, and also that a contact is only
possible when the languages are spoken at the same time.  We show how
computational methods of answer set programming and constraint logic
programming can be used to generate plausible conjectures about contacts
between prehistoric linguistic communities, and illustrate our approach by
applying it to the evolutionary history of Indo-European languages.
\end{abstract}


\section{Introduction} \label{sec:phylogeny-introduction}

The evolutionary history of families of natural languages is a major topic of
research in historical linguistics.  It is also of interest to archaeologists,
human geneticists, and physical anthropologists.  In this paper we show how
this work can benefit from the use of computational methods of logic
programming.

Our starting point here is the mathematical model of evolution of natural
languages introduced in \cite{rin02} and \cite{nak05}.  As proposed in
\cite{erd03}, we describe the evolution of languages in a declarative
formalism and generate conjectures about the evolution of Indo-European
languages using an answer set programming system.  Instead of the system
\hbox{\sc smodels},\footnote{\tt http://www.tcs.hut.fi/Software/smodels/ .}
employed in earlier experiments, we make use of the
new system \hbox{\sc
cmodels},\footnote{\tt http://www.cs.utexas.edu/users/tag/cmodels/ .}
which leads in this case to much better computation times.  Our main
conceptual contribution is extending the definition of a phylogenetic
network from \cite{nak05} by explicit temporal information about extinct
languages---by estimates of the dates when those languages could be spoken.
Computationally, to accommodate this enhancement we divide the work between
two systems: \hbox{\sc cmodels} and the constraint logic programming
system \hbox{\sc ecl}$^{\rs i}${\sc ps}$^{\rs e}$
({\tt http://www-icparc.doc.ic.ac.uk/eclipse/}).

\begin{figure}[t!]
\begin{center}
$\begin{array}{cc}
\epsfig{file=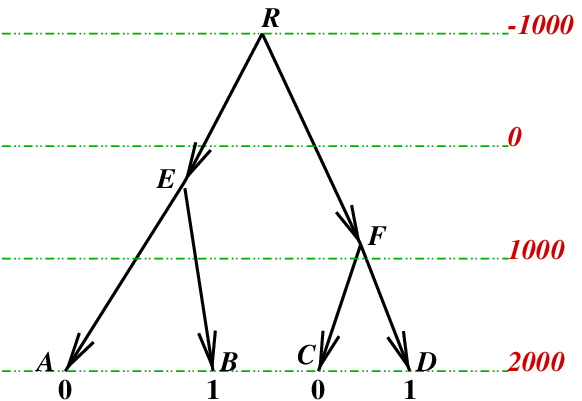} &
        \epsfig{file=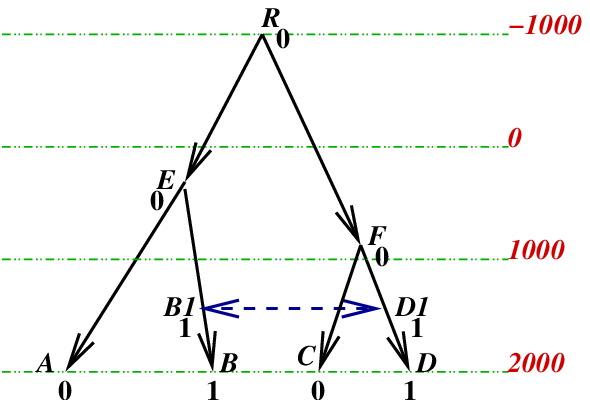} \ \\ [0.4cm]
\mbox{(a)} & \mbox{(b)}
\end{array}$
\end{center}
\caption{A temporal phylogeny (a), and a  perfect temporal network
(b) with a lateral edge connecting $B1$ with $D1$.}
\label{fig:phylogeny-small}
\end{figure}

It was observed long ago (see, for instance, \cite{gle59}) that if
linguistic communities do not remain in effective contact as their
languages diverge then the evolutionary history of their language family
can be modeled as a phylogeny---a tree whose edges represent genetic
relationships between languages.\footnote{We understand genetic 
relationships between languages in terms of linguistic ``descent'':
A language $Y$ of a given time 
is descended from a language $X$ of
       an earlier time if and only if $X$ developed into $Y$ by means of an
       unbroken sequence of instances of native-language
       acquisition by children.} 
Fig.~\ref{fig:phylogeny-small}(a)
shows the extant languages  $A$, $B$, $C$, $D$, along with the common
ancestor $E$
of $A$ and $B$, the common ancestor $F$ of $C$ and $D$, and the common
ancestor $R$
(for ``root") of~$E$ and~$F$.  The time line to the right of the tree shows
that the ancestors of $A$ and $B$ diverged around 300 CE,  the ancestors of
$C$ and $D$ diverged around 800 CE, and the ancestors of $E$ and $F$ diverged
around 1000 BCE.

Sometimes languages inherit their characteristics from their ancestors,
and sometimes they trade them with other languages when two linguistic
communities are in contact with each other.  The directed graph in
Fig.~\ref{fig:phylogeny-small}(b), obtained from the tree in
Fig.~\ref{fig:phylogeny-small}(a) by inserting two vertices and adding
a bidirectional edge, shows
that the ancestor $B1$ of $B$, spoken around 1400 CE, was in contact with
the ancestor $D1$ of $D$.

This idea has led Nakhleh, Ringe and Warnow~\citeyear{nak05}
to the definition of a phylogenetic
network, which extends the definition of a phylogeny by allowing lateral
edges, such as the edge connecting $B1$ with $D1$.\footnote{It was once 
customary to oppose a ``tree model'' of language
diversification, in which languages speciate relatively cleanly to
produce a definite phylogenetic tree, and a ``wave model'', in which
dialects evolve in contact, sharing innovations in overlapping
patterns which are inconsistent with a phylogenetic tree (see, e.g.,
\cite[pages 444-455]{hoc86} with references on page 667).  
But active researchers
have long recognized that each model is appropriate to a variety of
different real-world situations (cf. the discussion of 
Ross~\citeyear{ros97}). 
It therefore makes sense to explore models that incorporate the
strengths of both, such as tree models which incorporate edges
representing contact episodes between already diversified languages.}
The modification of their
mathematical model proposed below takes into account the fact
that two languages cannot be in contact unless they are spoken at the
same time. Geometrically speaking, every lateral edge has to be horizontal.
For instance, in Fig.~\ref{fig:phylogeny-small}(a) there can be no contact
between an ancestor
of~$E$ and a descendant of~$F$, although such contacts are not prohibited
in the definition of a phylogenetic network.  To express the idea of a
chronologically possible network in a precise form, we introduce ``temporal
networks"---networks with a ``date'' assigned to each vertex.
Dates monotonically increase along every branch of the phylogeny, and the
dates assigned to the ends of a lateral edge are equal to each
other.\footnote{These two assumptions imply the ``weak acyclicity" condition
from the definition of a phylogenetic network in \cite[Section 12.1]{nak04}.}

Once dates are assigned to the vertices of a phylogeny, we can talk not
only about the languages that are represented by the vertices, but also
about the ``intermediate" languages spoken by members of a linguistic
community at various times.  In the example above we would represent the
language spoken by the ancestors of the linguistic community $A$ at time $t$ by
the pair $A\up t$, where $300<t<2000$. This pair can be visualized as the point
at level $t$ on the edge leading to $A$.  In our idealized representation,
$t$ ranges over real numbers, so that the set of such pairs is infinite.
Language $B1$ in Fig.~\ref{fig:phylogeny-small}(b) can be denoted by
$B\up 1400$, and $D1$ can be written as $D\up 1400$.

The characteristics of a language that it could inherit from ancestors or
trade with other languages are called (qualitative) characters.
A phylogeny describes every leaf of the tree in terms of the values, or
``states,'' of the characters.   For instance, zeroes and
ones next to $A$, $B$, $C$ and $D$ in Fig.~\ref{fig:phylogeny-small}(a)
represent the states of a 2-valued character.  They can indicate, for example,
that a certain meaning is expressed by cognates\footnote{Note that we here
use the term 'cognates' as a  cover term for true cognates (inherited from
a common  ancestor) and words shared because of borrowing.  There does not
seem to be a convenient term that covers both types of cases.} in languages
$A$ and $C$ (cognation class 0), and that it is also expressed by cognates in
languages $B$ and $D$ (cognation class 1).

The main definition in this paper (similar to Definition 12.1.3 from
\cite{nak04})
is that of a ``perfect'' temporal network.  A perfect network explains how
every state of every qualitative character could evolve from its original
occurrence in a single language in the process of inheriting characteristics
along the tree edges and trading characteristics along the lateral edges of
the network.  For instance, Fig.~\ref{fig:phylogeny-small}(b) extends the
phylogeny in Fig.~\ref{fig:phylogeny-small}(a) to a perfect network by
labeling the internal vertices of the graph.  Indeed, consider the subgraph of
Fig.~\ref{fig:phylogeny-small}(b) induced by the set $\{A,C,E,F,R\}$ of
the vertices that are labeled 0.  This subgraph is a tree with the root~$R$;
this fact shows that state 0 has evolved in this network from its original
occurrence in $R$.
Similarly, the subgraph of Fig.~\ref{fig:phylogeny-small}(b)
induced by the set $\{B,B1,D,D1\}$ of the vertices labeled 1 contains a tree
with the root $B1$, and also a tree with the root $D1$.  This means that state
1 could evolve from its original occurrence in language $B1$ (or in an ancestor
of $B1$ that is younger than~$E$).  Alternatively, state 1 could evolve from
its original occurrence in language $D1$, or in an ancestor of $D1$ that is
younger than~$F$.

The computational problem that we are interested in is the problem of
reconstructing the temporal network representing the evolution of a
language family, such as Indo-European languages.  This problem can be
divided into two parts: generating a ``near-perfect'' phylogeny, and then
generating a small set of additional horizontal edges that turn the phylogeny
into a perfect network.  (In a plausible conjecture about the
evolution of languages the number of lateral edges has to be small:
inheriting characteristics of a language from its ancestors is far more
probable than acquiring them through borrowing, unless the dataset includes
a large proportion of words that are highly likely to be borrowed.)
The first part---generating phylogenies---has been the subject of extensive
research; applying answer set programming to this problem is discussed in
\cite{bro05}. In this paper we address the second part of the
problem---turning a phylogeny into a perfect network.

As to the dates assigned to the
vertices of the phylogeny, we assume that they are known approximately.
For instance, about the graph from Fig.~\ref{fig:phylogeny-small}(a)
we may only know that language $E$ was spoken between 100 BCE and
500 CE, and that language $F$ was spoken between 600 CE and 1100 CE.  Since
these two intervals do not overlap, we can conclude from these assumptions,
just as in the case of the specific dates assigned to $E$ and $F$ in
Fig.~\ref{fig:phylogeny-small}(a), that a
contact between an ancestor of $E$ and a descendant of $F$ would be impossible.
If, however, the given temporal intervals for $E$ and $F$ were wider then such
a conclusion might not be warranted, and a contact between an ancestor
of $E$ and a descendant of $F$ might become an acceptable conjecture.

Our method allows us to turn a phylogeny,
along with temporal intervals assigned to its vertices, into a perfect
network by adding a small number of lateral edges, or to determine that
this is impossible.

In this paper, after describing the problem mathematically in
Section~\ref{sec:phylogeny-pblm-description}, we show how it can be solved
using ideas of answer set programming and constraint logic programming
(Section~\ref{sec:computing}), and then apply our approach to the problem
of reconstructing the evolutionary history of Indo-European languages
(Section~\ref{sec:ie}).

\section{Problem Description} \label{sec:phylogeny-pblm-description}

In this section we show how the problem of computing perfect temporal
phylogenetic networks built on a given temporal phylogeny can be described
as a graph problem.

Recall that a {\em rooted tree} is a digraph with a vertex of in-degree 0,
called the {\em root}, such that every vertex different from the root has
in-degree 1 and is reachable from the root. In a rooted tree, a vertex of
out-degree 0 is called a {\em leaf}.

\subsection{Temporal Phylogenies} \label{ssec:temporal-phylogenies}

A {\em phylogeny} is a finite rooted tree $\langle V,E\rangle$ along
with two finite sets $I$ and $S$ and a function $f$ from $L\times I$ to $S$,
where $L$ is the set of leaves of the tree.  The elements of $I$ are usually
positive integers (``indices'') that represent, intuitively, qualitative
characters, and elements of $S$ are possible states of these characters.  The
function $f$ ``labels'' every leaf $v$---the extant or most recently spoken
language in one of the branches---by mapping every index $i$ to the state
$f(v,i)$ of the corresponding character in that language.

\begin{figure}[t!]
\begin{center}
$\begin{array}{cc}
\epsfig{file=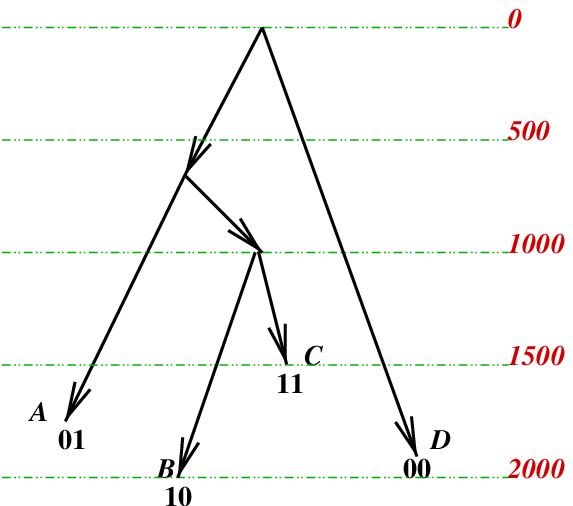} &
        \epsfig{file=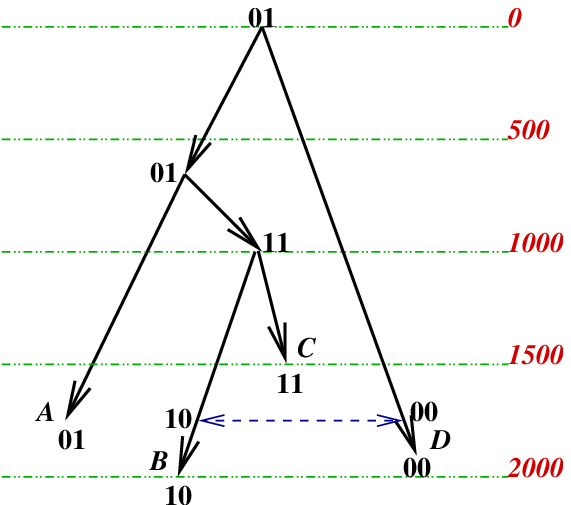} \\ [0.4cm]
\mbox{(a)} & \mbox{(b)}
\end{array}$
\end{center}
\caption{A temporal phylogeny (a), and a  perfect temporal network
(b) with a lateral edge connecting $B\up 1750$ with $D\up 1750$.}
\label{fig:phylogeny-t2}
\end{figure}

For instance, Fig.~\ref{fig:phylogeny-small}(a) is a phylogeny
with $I=\{1\}$ and $S=\{0,1\}$.
Fig.~\ref{fig:phylogeny-t2}(a) is a phylogeny with $I=\{1,2\}$
and $S=\{0,1\}$; state $f(v,i)$ is represented by the $i$-th member of the
tuple labeling leaf $v$.

A {\em temporal phylogeny} is a phylogeny along with a function $\tau$ from
vertices of the phylogeny to real numbers such that for every edge
$\langle u,v\rangle$ of the phylogeny $\tau(u)<\tau(v)$.  Intuitively,
$\tau(v)$ is the time when language $v$ was spoken.  We will graphically
represent the values of $\tau$ by placing a vertical time line to the right
of the tree, as in Fig.~\ref{fig:phylogeny-small}(a)
and Fig.~\ref{fig:phylogeny-t2}(a).

\subsection{Contacts and Networks} \label{ssec:contacts}

As discussed in the introduction, a contact between two linguistic communities
can be represented by a horizontal edge added to a pictorial representation
of a temporal phylogeny.  The two endpoints of the edge are simultaneous
``events'' in the histories of these communities.  An event can be represented
by a pair $v\up t$, where $v$ is a vertex of the phylogeny and $t$ is a real
number.

To make this idea precise, consider a temporal phylogeny $T$; let $V$ be the
set of its vertices, $R$ its root, and $\tau$ its time function.  For every
$v\in V\setminus \{R\}$, let
$\ii{par}(v)$ be the parent of $v$.  An {\em event} is any pair $v\up t$ such
that $v\in V\setminus\{R\}$ and $t$ is a real number satisfying the
inequalities
\beq
\tau(\ii{par}(v))<t\leq \tau(v).
\eeq{ineq}
Events $v\up t$ and $v'\up t'$ are {\em concurrent} if $t=t'$.  A {\em contact}
is a set consisting of two different concurrent events.

Any finite set $C$ of contacts defines a {\em (temporal phylogenetic)
network}---a digraph obtained from $T$ by inserting the elements $v\up t$ of
the contacts from $C$ as intermediate vertices and then adding every contact
in $C$ as a bidirectional edge.  We will discuss now a simple case that
is particularly important for applications, defined as follows.

We say that a set $C$ of contacts is {\em simple} if
\begin{itemize}
\item
for every event $v\up t$ that belongs to a contact from $C$, $t<\tau(v)$, and
\item
for every vertex $v$ of $T$ there exists at most one number $t$ such
that $v\up t$ belongs to some contact from $C$.
\end{itemize}
The first condition expresses that the second inequality in (\ref{ineq})
holds as a strict inequality, so that for every event $v\up t$ that belongs
to a contact from $C$
\beq
\tau(\ii{par}(v))<t< \tau(v).
\eeq{ineq1}
In other words, it says that the endpoints of all lateral
edges are different from the vertices of $T$; each of them subdivides an
edge of $T$ into two edges.  The second condition says that the endpoints of
the lateral edges do not subdivide any of the edges of $T$ into more than two
parts. It is clear, for instance, that the set consisting of the single contact
\beq
\{B\up 1400,D\up 1400\}
\eeq{c-example}
in Fig.~\ref{fig:phylogeny-small}
and the set consisting of the single contact
$$\{B\up 1750,D\up 1750\}$$
in Fig.~\ref{fig:phylogeny-t2} are simple.

If $C$ is simple then the corresponding network can be described as follows.
The set of its vertices is the union of the set $V$ of vertices of $T$ with
the union $V_C$ of the contacts from $C$.  Its set $E_C$ of edges is obtained
from the set $E$ of edges of $T$ in two steps.  First, for every event
$v\up t$ in $V_C$ we replace the edge $\langle\ii{par}(v),v\rangle$ from $E$
by its ``two halves''---the edges
$$\langle\ii{par}(v),v\up t\rangle\hbox{ and }\langle v\up t,v\rangle.$$
Second, for every contact $\{u\up t,v\up t\}$ in $ C$ we add a
``bidirectional lateral edge''---the pair of edges
$$\langle u\up t,v\up t\rangle\hbox{ and }\langle v\up t,u\up t\rangle.$$

\subsection{Perfect Networks} \label{ssec:perfect}

About a simple set $C$ of contacts (and about the corresponding network
$\langle V\cup V_C, E_C\rangle$) we say that it is {\em perfect} if there
exists a function
$g : (V\cup V_C)\times I \to S$ such that
\begin{enumerate}
\item[(i)]
    for every leaf $v$ of $T$ and every $i\in I$, $g(v,i)=f(v,i)$;
\item[(ii)]
    for every $i\in I$ and every $s\in S$, if the set
    $$V_{is} = \{x\in V\cup V_C\ :\ g(x,i)=s\}$$
    is nonempty then the digraph $\langle V\cup V_C,E_C\rangle$ has a subgraph
    with     the set $V_{is}$ of vertices that is a rooted tree.
\end{enumerate}
The first condition expresses that the function $g$ extends $f$ from leaves
to all ancestral languages of the network.  The
second condition expresses that every state $s$ of every character $i$ could
evolve from its original occurrence in some ``root'' language.

For instance, Fig.~\ref{fig:phylogeny-small}(b) shows a perfect network
obtained from the phylogeny of Fig.~\ref{fig:phylogeny-small}(a) by adding
one contact, along with labels representing the values of~$g$.  Similarly,
Fig.~\ref{fig:phylogeny-t2}(b) shows a perfect network obtained from the
phylogeny of Fig.~\ref{fig:phylogeny-t2}(a) along with the values of the
corresponding function $g$.  In the last example, state 0 of the first
character and state 1 of the second character have evolved from the
root of the given phylogeny; state 1 of the first character has evolved
from the common ancestor of $B$ and $C$; the state 0 of the second character
could evolve from $B\up 1750$ or from $D\up 1750$.
(Each of these two possibilities corresponds to a subgraph with the
vertices $B$, $D$, $B\up 1750$, $D\up 1750$ that is a rooted tree.)

\subsection{Increment to Perfect Simple Temporal Network}\label{ssec:ipstn}

We are interested in the problem of turning a temporal phylogeny into a
perfect temporal network by adding a small number of contacts.  For
instance, given the phylogeny in Fig.~\ref{fig:phylogeny-small}(a),
the single contact (\ref{c-example}) is a possible answer.

It is clear that the information included in a temporal phylogeny is not
sufficient for determining the exact dates of the contacts that turn it into
a perfect network.  For instance, if we shift contact (\ref{c-example}) up
or down by a few hundred years and replace it, say, by
$$\{B\up 1200,D\up 1200\}$$
then the new conjecture about the past of the languages $A,B,C,D$ will not be
distinguishable from~(\ref{c-example}).

To make this idea precise, let us select for each $v\in V\setminus\{R\}$
a new symbol $v\upp$, and define the {\em summary} of a simple set $C$ of
contacts to be the result of replacing each element $v\up t$ of every contact
in $C$ with $v\upp$.  Thus summaries consist of 2-element subsets of the set
$$V\upp=\{ v\upp\ :\ v\in V\setminus\{R\}\}.$$
For instance, the summary of the set of contacts
of Fig.~\ref{fig:phylogeny-small}(b) is $\{\{B\upp,D\upp\}\}$.
For the set of contacts of Fig.~\ref{fig:phylogeny-t2}(b), the summary is
the same.  Intuitively, $v\upp$ is a language intermediate between
$\ii{par}(v)$ and $v$ that was spoken at some unspecified time between
$\tau(\ii{par}(v))$ and $\tau(v)$.

An {\em IPSTN problem} (for ``Increment to Perfect Simple Temporal Network'')
is defined by a phylogeny $\langle V,E,I,S,f\rangle$ and a function
$$v\mapsto (\tau_{\is{min}}(v),\tau_{\is{max}}(v))$$
from the vertices of the phylogeny to open intervals.  (In other words,
for every $v\in V$, $\tau_{\is{min}}(v)$ is a real number or $-\infty$,
and $\tau_{\is{max}}(v)$ is a real number or $+\infty$, such that
$\tau_{\is{min}}(v)<\tau_{\is{max}}(v)$.)  A {\sl solution}
to the problem is a set of 2-element subsets of $V\upp$ that is the summary
of a perfect simple set of contacts for a temporal phylogeny
$\langle V,E,I,S,f,\tau\rangle$ such that,
for all $v\in V$,
\beq
 \tau_{\is{min}}(v) < \tau(v) < \tau_{\is{max}}(v).
\eeq{time}

Thus a solution is a summary, rather than a set of contacts itself.  On
the other hand, as discussed in the introduction, an IPSTN problem includes
a set of conditions on a time function, rather than a specific temporal
phylogeny.

Given an IPSTN problem $Q$ and a nonnegative integer $k$, we want to find the
solutions $X$ to $Q$ such that the cardinality of $X$ is at most $k$.

\section{Computing Solutions} \label{sec:computing}

Our approach to computing solutions is based on their characterization in
terms of ``admissible sets.'' Whether or not a set $X$ is admissible for
an IPSTN problem $Q$ is completely determined by the phylogeny of $Q$;
this property does not depend on the time intervals
$(\tau_{\is{min}}(v),\tau_{\is{max}}(v))$.  The problem of computing
admissible sets lends itself well to the use of answer set programming in
the spirit of \cite{erd03}.  Proposition~\ref{prop:quasi} below shows, on
the other hand, that solutions to $Q$ can be described as the admissible
sets for which a certain system of equations and inequalities has a
solution.  This additional condition is easy to verify, for each admissible
set, using a constraint programming system.

\subsection{Solutions as Admissible Sets} \label{ssec:admissible}

Consider a phylogeny $\langle V,E,I,S,f\rangle$ with a root $R$, and a set
$X$ of 2-element subsets of $V\upp$.  By $V_X$ we denote the union of all
elements of $X$.  By $E_X$ we denote the set obtained from $E$ by replacing,
for every $v\upp\in V_X$, the edge $\langle\ii{par}(v),v\rangle$ with
$$\langle \ii{par}(v),v\upp\rangle\hbox{ and }\langle v\upp,v\rangle$$
and adding, for every element $\{u\upp,v\upp\}$ of $X$, the edges
$$\langle u\upp,v\upp\rangle\hbox{ and }\langle v\upp,u\upp\rangle.$$

We say that $X$ is {\sl admissible} if there exists a function
$g: (V\cup V_X) \times I \to S$ such that
\begin{enumerate}
\item[(i)]
     for every leaf $v$ of the phylogeny and every $i\in I$,
     $g(v,i)=f(v,i)$;
\item[(ii)]
     for every $i\in I$ and every $s\in S$, if the set
     $$V_{is} = \{x\in V\cup V_X\ :\ g(x,i)=s\}$$
     is nonempty then the digraph $\langle V\cup V_X,E_X\rangle$
     has a subgraph with the set $V_{is}$ of vertices that is a rooted tree.
\end{enumerate}

In the following proposition, $Q$ is an IPSTN problem defined by a
phylogeny $\langle V,E,I,S,f\rangle$ with a root $R$ and a function
$v\mapsto (\tau_{\is{min}}(v),\tau_{\is{max}}(v))$.

\begin{prop}\label{prop:quasi}
A set $X$ of 2-element subsets of $V\upp$ is a solution to $Q$ iff
\begin{enumerate}
\item[(i)]
     $X$ is admissible, and
\item[(ii)]
there exists a real-valued function $\tau$ on $V\cup V_X$ such that
  \begin{enumerate}
  \item[(a)] for every $v\in V$,
            $$\tau_{\is{min}}(v)<\tau(v)<\tau_{\is{max}}(v),$$
  \item[(b)] for every $v\in V\setminus\{R\}$,
            $$\tau(\ii{par}(v))<\tau(v),$$
  \item[(c)] for every element $v\upp$ of $V_X$,
            $$\tau(\ii{par}(v))<\tau(v\upp)<\tau(v),$$
  \item[(d)] for every element $\{u\upp,v\upp\}$ of $X$,
            $$\tau(u\upp)=\tau(v\upp).$$
   \end{enumerate}
\end{enumerate}
\end{prop}

\noindent{\bf Proof}$\qquad${\sl Left-to-right.}  Assume that $X$ is a
solution to $Q$, so that there exist a real-valued function $\tau$ on $V$
satisfying (\ref{time}) and a perfect simple set $C$ of contacts for the
temporal phylogeny $\langle V,E,I,S,f,\tau\rangle$ such that $X$ is the
summary of $C$.  The function from $V_C$ to $V_X$ that maps every event
$v\up t$ to $v\upp$ is a 1--1 correspondence between the two sets.  If we
agree to identify every event $v\up t$ with its image $v\upp$ under this
correspondence then $E_C$ becomes identical to $E_X$, and the conditions on
$g$ in the definition of a perfect set of contacts turn into the conditions
on $g$ in the definition of an admissible set.  Consequently (i) follows from
the fact that $C$ is perfect.  To prove (ii), extend $\tau$ from $V$ to
$V\cup V_X$:
$$\tau(v\upp)=t\ \hbox{ if }\ v\up t \in V_C.$$
Part~(a) follows from~(\ref{time}); part~(b) follows from the definition of
a temporal phylogeny; part~(c) follows from~(\ref{ineq1});
part~(d) follows from the definition of a contact.

{\sl Right-to-left.} Assume that $X$ satisfies
conditions (i) and (ii).  Consider the temporal phylogeny $T$ that consists
of the given phylogeny $\langle V,E,I,S,f\rangle$ and the function $\tau$
restricted to $V$.  By~(a), $T$ satisfies~(\ref{time}).  Let $C$ be the set
obtained from $X$ by replacing each symbol $v\upp$ in every element of $X$
with the event $v\up t$ where $t=\tau(v\upp)$.  From~(d) we conclude that the
elements of $C$ are contacts; by~(c), $C$ is simple.  It is clear that $X$ is
the summary of $C$.  The same reasoning as in the first half of the proof
shows that, in view of~(i), $C$ is perfect.  

\subsection{Answer Set Programming}\label{ssec:asp}

The idea of answer set programming is to represent
a given computational problem as a logic program whose answer sets (stable
models) \cite{gel88} correspond to solutions.  Systems
that compute answer sets for a logic program are called answer set solvers.
System \hbox{\sc smodels} with its front-end~\hbox{\sc lparse}
is one of the most widely used answer set solvers today.  The system
\hbox{\sc cmodels} is another answer set solver, and it uses \hbox{\sc lparse}
as its front-end also.  This newer system does not have its own search
engine; it is essentially a compiler that translates the problem of
computing answer sets into a propositional
satisfiability problem (or into a series of propositional satisfiability
problems), and invokes a SAT solver, such as {\sc
zchaff},\footnote{\tt http://www.ee.princeton.edu/$\sim$chaff/zchaff.php .}
to perform search.

Unlike Prolog systems, which expect from the user a program and a query,
an answer set solver starts computing given a program only.  For
instance, we can give \hbox{\sc smodels} the input program
\begin{verbatim}
p(0).
q(1).
r(X) :- p(X).
r(X) :- q(X).
\end{verbatim}
and it will produce the output
\begin{verbatim}
Stable Model: r(0) r(1) q(1) p(0)
\end{verbatim}
---the set of all ground queries to which Prolog would respond {\tt yes}.
Given the program

\begin{verbatim}
p(0).
p(1).
q(1-X) :- p(X), not q(X).
\end{verbatim}

\noindent\hbox{\sc smodels} responds

\begin{verbatim}
Answer: 1
Stable Model: q(1) p(1) p(0) 
Answer: 2
Stable Model: q(0) p(1) p(0) 
\end{verbatim}

\noindent
This output means, intuitively, that the program can be understood in two
ways: either {\tt q(0)} is false and {\tt q(1)} is true, or the other way
around.  For this program (and for other programs with several answer sets)
there is no simple relationship between the behavior of Prolog and the
behavior of answer set solvers.

A Prolog program can be viewed as a collection of definitions of predicates.
In addition to such ``defining'' rules, \hbox{\sc lparse} programs often
include rules of two other kinds--- ``choice rules'' and ``constraints.''
For example,
\begin{verbatim}
{p,q,r,s}.
\end{verbatim}
is a choice rule.  Its answer sets are arbitrary subsets of
$\{{\tt p},{\tt q},{\tt r},{\tt s}\}$.  Intuitively, this rule says: for
each of the atoms {\tt p}, {\tt q}, {\tt r}, {\tt s}, choose arbitrarily
whether to include it in the answer set.  A choice rule may include
restrictions on the cardinality of the answer set.  For instance, the
answer sets of
\begin{verbatim}
2 {p,q,r,s} 3.
\end{verbatim}
are the subsets $\{{\tt p},{\tt q},{\tt r},{\tt s}\}$ whose cardinality is
at least 2 and at most 3.

A constraint is, syntactically, a rule with the empty head.  The effect of
adding a constraint to a program is to make the collection of its answer
sets smaller---to remove the answer sets that ``violate'' the constraint.
For instance, by adding the constraint
\begin{verbatim}
:- p, not q.
\end{verbatim}
to a program we remove its answer sets that include {\tt p} and do not include
{\tt q}.

A detailed description of the input language of \hbox{\sc lparse} can be
found in the online manual ({\tt
http://www.tcs.hut.fi/Software/smodels/lparse.ps.gz}).

\subsection{Generating Admissible Sets}
\label{ssec:generating}

An {\sc lparse} program for generating admissible sets is shown in
Fig.~\ref{fig:lparse}.
\begin{figure}[t!]
\begin{verbatim}
#domain vertex(U;V).
#domain character(I).
#domain state(S).

{x(UU,VV): vertex(UU;VV): UU != r: VV != r: UU < VV} k.

xx(U,V) :- x(U,V), U < V.
xx(V,U) :- x(U,V), U < V.

v_x(U) :- xx(U,V).

% definition of admissibility, part (i)

g(V,I,S) :- f(V,I,S).
1 {g(V,I,SS): state(SS)} 1 :- e(V,U).
1 {g(pre(V),I,SS): state(SS)} 1 :- v_x(V).

% definition of admissibility, part (ii)

{root(V,I,S)} :- g(V,I,S).
{root(pre(V),I,S)} :- g(pre(V),I,S).

:- root(U,I,S), root(V,I,S), U < V.
:- root(U,I,S), root(pre(V),I,S).
:- root(pre(U),I,S), root(pre(V),I,S), U < V.

reachable(V,I,S) :- root(V,I,S).
reachable(pre(V),I,S) :- root(pre(V),I,S).
reachable(pre(V),I,S) :- e(U,V), g(pre(V),I,S), reachable(U,I,S).
reachable(V,I,S) :- v_x(V), g(V,I,S), reachable(pre(V),I,S).
reachable(V,I,S) :- e(U,V), not v_x(V), g(V,I,S), reachable(U,I,S).
reachable(pre(U),I,S) :- xx(U,V), g(pre(U),I,S), reachable(pre(V),I,S).

:- g(V,I,S), not reachable(V,I,S).
:- g(pre(V),I,S), not reachable(pre(V),I,S).
\end{verbatim}
\caption{An {\sc lparse} program for generating admissible sets.}
\label{fig:lparse}
\end{figure}
Table~\ref{table:explanation} shows the correspondence between the symbols
used in the program and the notation introduced in
Sections~\ref{sec:phylogeny-pblm-description}, ~\ref{ssec:admissible}
and~\ref{ssec:generating}.\footnote{The representation of ${\tt V}\upp$ by
{\tt pre(V)} is suggested by the distinction between a ``proto'' language
and its ``pre-proto'' stage in historical linguistics.  The term
``proto-Germanic,'' for instance, represents a language that was about to
split up into Germanic languages,
each spoken by a different speech community; the speech of the ancestors
of the proto-Germanic generation, slowly changing all the time, is
referred to as pre-proto-Germanic.} The program should be combined with
the definition of the domain predicates {\tt vertex}, {\tt e}, {\tt character},
{\tt state}, {\tt f} describing the given phylogeny. 

The first three lines of the program tell \hbox{\sc lparse} that {\tt U} and
{\tt V} range over vertices, {\tt I} ranges over characters, and {\tt S} over
states.
The vertices of the phylogeny are assumed to be integers, and the expression
\hbox{\tt U < V} in the program is understood accordingly.
The verification of condition (ii) from the definition of an admissible set
is based on the fact that (ii) can be equivalently stated as follows:
for every $i\in I$ and every $s\in S$, if the set $V_{is}$ is nonempty then
there is a vertex $v_{is}\in V_{is}$ such all elements of $V_{is}$ are
reachable from $v_{is}$ in $V_{is}$ \cite[Proposition~1]{erd03}.

\begin{table}
\caption{Explanation of the symbols used in Fig.~\ref{fig:lparse}.}
\label{table:explanation}
\begin{tabular}{ll}
\hline\hline
{\sc lparse} program & Mathematical notation\\
\hline
{\tt vertex(V)}        & ${\tt V}\in V$                        \\
{\tt character(I)}     & ${\tt I}\in I$                        \\
{\tt state(S)}         & ${\tt S}\in S$                        \\
{\tt e(U,V)}           & $\langle{\tt U},{\tt V}\rangle\in E$  \\
{\tt r}                & $R$                                   \\
{\tt pre(V)}           & ${\tt V}\upp$                         \\
{\tt x(U,V)}           & $\{{\tt U}\upp,{\tt V}\upp\}\in X$ and
                         ${\tt U}<{\tt V}$                     \\
{\tt xx(U,V)}          & $\{{\tt U}\upp,{\tt V}\upp\}\in X$    \\
{\tt v\_x(V)}          & ${\tt V}\upp\in V_X$                  \\
{\tt f(V,I,S)}         & $f({\tt V},{\tt I})={\tt S}$          \\
{\tt g(V,I,S)}         & $g({\tt V},{\tt I})={\tt S}$          \\
{\tt root(V,I,S)}      & ${\tt V}=v_{\tt IS}$                  \\
{\tt reachable(V,I,S)}
$\qquad$               & {\tt V} is reachable from $v_{\tt IS}$
                         in $V_{\tt IS}$\qquad$$               \\
\hline\hline
\end{tabular}
\end{table}

The algorithm for solving IPSTN problems suggested by the discussion above
consists of two steps: compute admissible sets by running an answer set
solver on the program from Fig.~\ref{fig:lparse} and then use a constraint
programming system to check, for each of these sets $X$, whether the equations
and inequalities from part~(ii) of the statement of
Proposition~\ref{prop:quasi} have a solution in real numbers $\tau(v)$,
$v\in V\cup V_X$.

This basic algorithm can be improved using the following observation.
Let $X$ be a solution to the given IPSTN problem.  Consider the numbers
$\tau(v)$ from part (ii) of the statement of
Proposition~\ref{prop:quasi}.  Conditions (a) and (c) imply that for every
$v\upp$ in $V_X$
$$\tau_{\is{min}}(\ii{par}(v))<\tau(\ii{par}(v))<\tau(v\upp)<
                                     \tau(v)<\tau_{\is{max}}(v),$$
so that $v\upp$ belongs to the interval
$(\tau_{\is{min}}(\ii{par}(v)),\tau_{\is{max}}(v))$.
In view of (d), it follows that for every element $\{u\upp,v\upp\}$ of $X$,
the intervals
$$
(\tau_{\is{min}}(\ii{par}(u)),\tau_{\is{max}}(u))\hbox{ and }
(\tau_{\is{min}}(\ii{par}(v)),\tau_{\is{max}}(v))
$$
overlap.  Consequently, extending the program from Fig.~\ref{fig:lparse}
by the constraints
\beq
\hbox{\tt :- x($u,v$).}
\eeq{constraint}
for the pairs $u$, $v$ for which these intervals do {\sl not} overlap
will not lead to the loss of solutions.

\subsection{Making the Program Tight}

The operation of \hbox{\sc cmodels} \cite{giu04} is based on the fact that
the answer sets for a program can be described as the models of the program's
completion that satisfy its loop formulas \cite{lin02}.  This process is
particularly simple in the case when the program is tight, because a tight
program has no loops, and its set of loop formulas is empty \cite{erd03a},
\cite[Section~5]{lee03a}.  The difference between tight and non-tight programs
can be illustrated with a simple example: a program containing the rules
\begin{verbatim}
p :- q,r.
q :- p,s.
\end{verbatim}
is not tight, because it contains the loop $\{{\tt p},{\tt q}\}$.

The usual recursive definition of the reachability of a vertex in a digraph
is tight only when the graph is acyclic.  In Fig.~\ref{fig:lparse}, the
atom {\tt reachable(V,I,S)}  expresses that {\tt V} can be reached from
$v_{\tt IS}$ in the subgraph of the network induced by $V_{\tt IS}$; since the
network contains cycles, the program in Fig.~\ref{fig:lparse} is not tight.

But we can make this program tight using a transformation somewhat
similar to the process of tightening described in \cite[Section~3.2]{lif96b}.
The network is obtained from a tree by adding at most {\tt k}
bidirectional lateral edges.  Consequently, the shortest path between any pair
of vertices in the network includes at most {\tt k} lateral edges.    Consider
the auxiliary atoms {\tt rj(V,I,S,J)}, expressing that there exists a path
from $v_{\tt IS}$ to {\tt V} in $V_{\tt IS}$ that contains exactly {\tt J}
lateral edges ($0\leq{\tt J}\leq{\tt k}$).  The predicate~{\tt rj} can be
characterized by a tight definition:

\begin{verbatim}
rj(V,I,S,0) :- root(V,I,S).
rj(pre(V),I,S,J) :- root(pre(V),I,S).
rj(pre(V),I,S,J) :- e(U,V), g(pre(V),I,S), rj(U,I,S,J).
rj(V,I,S,J) :- v_x(V), g(V,I,S), rj(pre(V),I,S,J).
rj(V,I,S,J) :- e(U,V), not v_x(V), g(V,I,S), rj(U,I,S,J).
rj(pre(U),I,S,J+1) :- xx(U,V), g(pre(U),I,S), rj(pre(V),I,S,J), J < k.
\end{verbatim}
Then {\tt reachable} can be defined in terms of {\tt rj} by the rules
\begin{verbatim}
reachable(V,I,S) :- rj(V,I,S,J).
reachable(pre(V),I,S) :- rj(pre(V),I,S,J).
\end{verbatim}

This optimization has a significant effect on the computation time of
\hbox{\sc cmodels}.

\section{Contacts Between Indo-European Languages}\label{sec:ie}

We have applied the concept of a temporal phylogenetic network and the
computational methods described above to the problem of generating
conjectures about contacts between prehistoric Indo-European languages,
discussed earlier in \cite{nak05}, and also in \cite{erd03} and
\cite[Chapter 13]{nak04}.
In these experiments, \hbox{\sc cmodels} was used as the answer set solver, and
\hbox{\sc ecl}$^{\rs i}${\sc ps}$^{\rs e}$ as the constraint programming
system.

The problems addressed in these experiments are more general than IPSTN
problems discussed in Sections~\ref{sec:phylogeny-pblm-description}
and~\ref{sec:computing}: we were interested in
sets of contacts that are not necessarily simple in the sense of
Section~\ref{ssec:contacts}.  The theory and the computational methods
presented above have been extended to ``IPTN problems'' involving the
networks that may have several lateral edges meeting the same tree edge,
and that may have lateral edges incident to the vertices of the given
phylogeny---the possibilities ruled out in the definition of a simple set
of contacts.
The program shown in Fig.~\ref{fig:lparse} has been modified accordingly, and
it was also optimized
by allowing function $g$ to be partial, as in \cite[Section~5]{erd03}.

\subsection{A Phylogeny of Indo-European Languages}\label{ssec:ie}

As the starting point, we took a phylogeny of Indo-European languages
based on the ``unscreened IE dataset'' published at {\tt
http://www.cs.rice.edu/$\sim$nakhleh/CPHL/} (without characters that are
uninformative or that exhibit known parallel development of states),
and on the genetic tree shown in
\cite[Fig.~5]{nak05} (published originally in \cite{rin02}).
Using the methods discussed in \cite[Sections~3 and~4]{erd03}, we
extracted from that phylogeny a small part that appears to
contain all components essential for the task of reconstructing contacts
between prehistoric Indo-European languages.

\begin{figure}[t!]
\begin{center}
    \epsfig{file=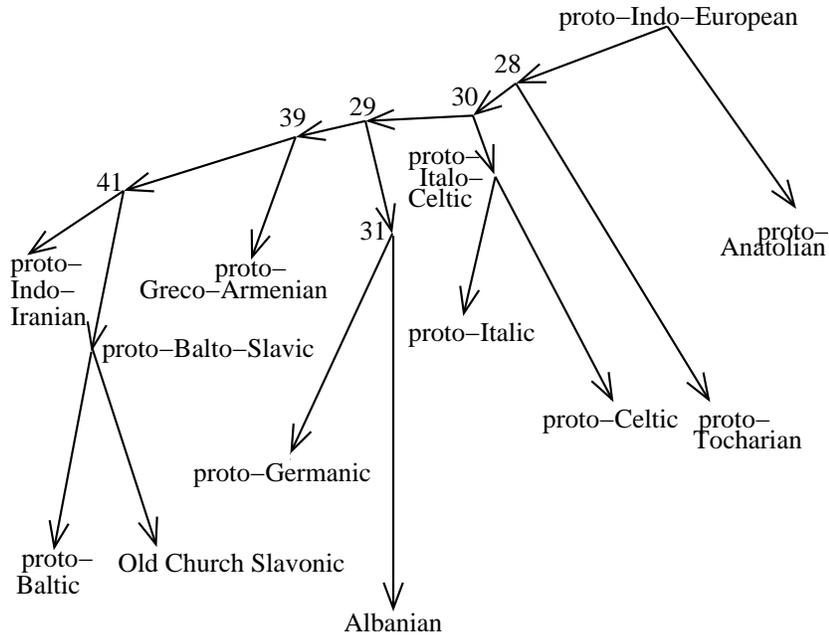,scale=0.9}
\end{center}
\caption{A phylogenetic tree of Indo-European languages.  The languages
that do not have commonly accepted names are labeled by numbers.}
\label{fig:pruned-tree}
\end{figure}

The vertices and edges of this smaller phylogeny are shown in
Fig.~\ref{fig:pruned-tree}.
All vertices except two (Old Church Slavonic
and Albanian) are ``prehistoric'' languages, reconstructed by comparing their
descendants.  For instance, proto-Celtic has been reconstructed from what is
known about its recorded descendants, Old Irish and Welsh (and the
fragmentarily attested Continental Celtic  languages of antiquity).
The phylogeny has 16 qualitative characters (all lexical), and each 
character has 2 or 3
states. Some of the essential character states are shown in 
Table~\ref{tab:ie}.\footnote{Let $\langle V,E\rangle$ be a phylogeny along
with a set $I$ of characters, a set $S$ of character states, 
and a function $f$ from $L\times I$ to $S$,
where $L$ is the set of leaves of the tree. A state $s\in
S$ is {\em essential} with respect to a character $j\in I$ if there exist two
different leaves $l_1$ and $l_2$ in $L$ such that $f(l_1,j) = f(l_2,j) = s$.}

\begin{table}[t!]
\caption{The essential character states of some lexical characters 
for the languages denoted by the leaves 
of the phylogeny in Fig.~\ref{fig:pruned-tree}.}
\label{tab:ie}
\begin{tabular}{lcccccc}
\hline\hline
                    & `one' & `arm' & `beard' & `free' & `pour' & `tear' \\
\hline
proto-Indo-Iranian  &      &  5     &  1      &       &        &  11 \\
proto-Baltic        &  11  &  8     &  5      &       &  6     &  11 \\
Old Church Slavonic &      &        &  5      &       &  6     &    \\
proto-Greco-Armenian&  2   &        &  1      &  3    &  3     &  2 \\
proto-Germanic      &  11  &  8     &  5      &  10   &  14    &  2 \\
Albanian            &  2   &        &  1      &       &        &   \\
proto-Italic        &  11  &        &  5      &  3    &  14    &  2 \\
proto-Celtic        &  11  &        &         &  10   &        &  2 \\
proto-Tocharian     &  2   &   5    &         &       &   3    &  11 \\  
proto-Anatolian     &      &        &  1      &       &        &     \\
\hline\hline
\end{tabular}
\end{table}

Table~\ref{table:intervals} shows, for each vertex of the tree, our
assumptions about the time when the corresponding language was
spoken.  Our calculations assume, for instance, that proto-Indo-Iranian was
spoken by a generation that lived between 2100~BCE and 1700~BCE.

\begin{table}[t!]
\caption{Time intervals for the languages from Fig.~\ref{fig:pruned-tree}.}
\label{table:intervals}
\begin{tabular}{lrl}
\hline\hline
\qquad\qquad $\;\;v$   &$(\tau_{\is{min}}(v)$,&$\!\!\!\! \tau_{\is{max}}(v))$\\
\hline
proto-Indo-European           &  $(-4500$,&$\!\!\!\! -3800)$\\
proto-Indo-Iranian            &  $(-2100$,&$\!\!\!\! -1700)$\\
proto-Balto-Slavic            &  $(-1400$,&$\!\!\!\! -800)$\\
proto-Baltic                  &  $(600$,&$\!\!\!\! 1000)$\\
Old Church Slavonic           &   $(870$,&$\!\!\!\! 1000)$\\
proto-Greco-Armenian          &  $(-2500$,&$\!\!\!\! -2200)$\\
proto-Germanic                &  $(-400$,&$\!\!\!\! 0)$\\
Albanian                      &   $(1800$,&$\!\!\!\! 2100)$\\
proto-Italo-Celtic            &  $(-3000$,&$\!\!\!\! -2400)$\\
proto-Italic                  &  $(-1500$,&$\!\!\!\! -1000)$\\
proto-Celtic                  &  $(-700$,&$\!\!\!\! -300)$\\
proto-Tocharian               &  $(-700$,&$\!\!\!\! -300)$\\
proto-Anatolian               &  $(-2500$,&$\!\!\!\! -2100)$\\
Vertex 28                     &  $(-3900$,&$\!\!\!\! -3300)$\\
Vertex 29                     &  $(-3600$,&$\!\!\!\! -3000)$\\
Vertex 30                     &  $(-3500$,&$\!\!\!\! -2900)$\\
Vertex 31                     &  $(-2400$,&$\!\!\!\! -1800)$\\
Vertex 39                     &  $(-3400$,&$\!\!\!\! -2800)$\\
Vertex 41                     &  $(-2600$,&$\!\!\!\! -2200)$\\
\hline\hline
\end{tabular}
\end{table}

Estimating the dates of prehistoric languages is a matter of informed 
guesswork, because rates of linguistic change are known to vary not 
only over time but also between lineages (see especially \cite{ber62}).
Relevant archaeological evidence must be taken 
into account, but it rarely settles important disputes, because the 
material remains of a culture typically reveal nothing about the 
language (or languages) spoken, in the absence of written documents.  
The dates suggested here for internal nodes of the IE tree are
estimates and are presented with considerable diffidence.  For a good 
summary and discussion of the archaeological evidence the reader is 
referred to \cite{mal89}.   

Some solutions in the sense of Section~\ref{sec:phylogeny-pblm-description}
do not represent viable conjectures about the evolution of Indo-European
languages for geographical reasons.  For
instance, a contact between pre-proto-Celtic and pre-proto-Baltic
is unlikely because the former was spoken in western Europe, while
the Balts were probably confined to a fairly small area in
northeastern Europe.  We have eliminated several unrealistic possibilities
of this kind at the stage of computing admissible sets, by including
additional constraints of the form~(\ref{constraint}). 
For instance, the possibility above can be eliminated by adding to 
the {\sc lparse} program that generates admissible sets the constraint:
\begin{verbatim}
:- x(38,43).
\end{verbatim}
where {\tt 38} denotes proto-Celtic and {\tt 43} denotes proto-Baltic.

\subsection{Results}\label{ssec:results}

\begin{figure}[t!]
  \begin{center}
      \epsfig{file=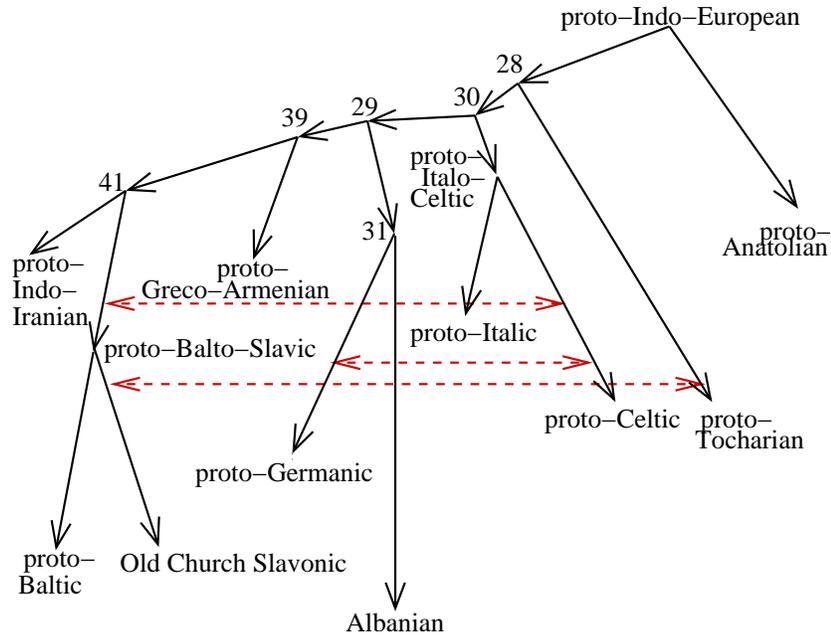,scale=0.9}
  \end{center}
  \caption{A conjecture about contacts between Indo-European languages
  generated by {\sc cmodels} and accepted by
  {\sc ecl}$^{\rs i}${\sc ps}$^{\rs e}$.}
  \label{fig:ie-contacts}
 \end{figure}

The problem described in Section~\ref{ssec:ie}, with the additional
geographical constraints mentioned above,
turns out to have no solutions consisting of fewer than 3 contacts.
There are three solutions of cardinality 3. (To be precise,
we should say ``three essentially different solutions,'' because a summary
does not specify the exact times of contacts.)
The first (Fig.~\ref{fig:ie-contacts}) involves contacts between

\medskip
\begin{tabular}{l}
pre-Old Church Slavonic and pre-proto-Tocharian,\\
pre-proto-Germanic and pre-proto-Celtic,\\
pre-proto-Balto-Slavic and pre-proto-Celtic;
\end{tabular}
\medskip

\noindent
the second, contacts between

\medskip
\begin{tabular}{l}
pre-Old Church Slavonic and pre-proto-Tocharian,\\
pre-proto-Germanic and pre-proto-Italic,\\
pre-proto-Italic and pre-proto-Balto-Slavic;
\end{tabular}
\medskip

\noindent
the third, contacts between

\medskip
\begin{tabular}{l}
pre-proto-Italic and pre-proto-Greco-Armenian,\\
pre-proto-Germanic and pre-proto-Italic,\\
pre-proto-Baltic and pre-proto-Germanic.
\end{tabular}
\medskip

\noindent
All three summaries generated by {\sc cmodels} have been accepted by the
\hbox{\sc ecl}$^{\rs i}${\sc ps}$^{\rs e}$ filter as solutions (which means
that all relevant chronological information was expressed in this case by the
constraints shown at the end of Section~\ref{ssec:generating}).
They have been computed in about 40 minutes of CPU
time using {\sc lparse~1.0.13}, {\sc cmodels 2.10},
{\sc zchaff Z2003.11.04}, and \hbox{\sc ecl}$^{\rs i}${\sc ps}$^{\rs e}$
{\sc 3.5.2}, on a PC with a 733 Intel Pentium III processor and 256MB RAM,
running SuSE Linux (Version~8.1).

We have also determined, using {\sc cmodels}, that there exist 193 admissible
sets of cardinality 4 that are minimal with respect to set inclusion; out of
those, 14 have been rejected by \hbox{\sc ecl}$^{\rs i}${\sc ps}$^{\rs e}$.
Some of the 4-edge solutions represent plausible
conjectures about the history of Indo-European languages.  One such solution
includes, for instance, contacts between

\medskip
\begin{tabular}{l}
pre-Old Church Slavonic and pre-proto-Tocharian,\\
pre-proto-Germanic and pre-proto-Italic,\\
pre-proto-Germanic and pre-proto-Celtic,\\
pre-proto-Germanic and pre-proto-Baltic.
\end{tabular}

\subsection{Comparison with Earlier Work}

The three 3-edge solutions listed in Section~\ref{ssec:results}
are identical to the solutions that
are marked as ``feasible'' in \cite[Table~3]{nak05}.  That table shows the 16
sets of lateral edges generated by MIPPN, the software tool designed for
solving the Minimum Increment to Perfect Phylogenetic Network problem.  It
is different from the computational problem that we solve here using logic
programming tools in that its input does not include any chronological or
geographical information.  The 16 sets of contacts produced by MIPPN were
scrutinized by a specialist in the history of Indo-European languages, who
has determined that most of them are not plausible from the point of view
of historical linguistics.  Then the remaining 3 sets were declared feasible.
The logic programming approach, on the other hand, allowed us to express the
necessary expert knowledge about chronological and geographical constraints
in formal notation, and to give this information to the program as
part of input, along with the phylogeny.  All ``implausible'' solutions
were weeded out in this case by \hbox{\sc cmodels} without human intervention.

In the experiments described in \cite{erd03}, chronological and geographical
information was not part of the input either.  But those experiments were
similar to the work described in this paper in that search, in both cases,
was performed
using answer set solvers: {\sc smodels} in \cite{erd03}, and {\sc cmodels}
with {\sc zchaff} in this project.  The difference in the computational
efficiency between the two engines turned out to be significant.
With the new tools available, we did not have to employ
the ``divide-and-conquer'' strategy described in \cite[Section~6]{erd03}.
The time needed to compute the 3-edge solutions went down from over 150 hours
to around 40 minutes.  For comparison, the computation time of MIPPN in the
same application was around 8 hours \cite[Section~5.3]{nak05}.

\section{Conclusion}

The mathematical model of the evolutionary history of natural languages
proposed in \cite{nak05} enriched the traditional ``evolutionary tree''
model by allowing languages in different branches of the tree to trade their
characteristics.  In that theory, phylogenetic networks take place of trees.
In this paper we discussed a further enhancement of the phylogenetic network
model, which incorporates a real-valued function assigning times to the
vertices of the network and prohibits a contact between two languages
if it is chronologically impossible.  The use of the time function allows
us to reduce the number of networks that are mathematically ``perfect'' but
do not represent historically plausible conjectures.

Computing perfect temporal networks can be
accomplished by a combination of an answer set programming ``generator'' with
a constraint logic programming ``filter.''  An alternative approach to
combining computational methods developed in these two subareas of logic
programming is discussed in \cite{elk04}.                               

In application to the problem of computing perfect networks for a
phylogeny of Indo-European languages, the use of {\sc cmodels} with
{\sc zchaff} has improved the computation time by two orders of
magnitude in comparison with the use of {\sc smodels} in earlier
experiments.

\section*{Acknowledgments}
We are grateful to Michael Gelfond and the anonymous referees for 
useful suggestions. Vladimir Lifschitz was partially supported by 
the National Science Foundation under Grant IIS-0412907. 
Don Ringe was supported by the National Science Foundation under 
Grant ITR-0321911.
Esra Erdem was supported in part by the Austrian Science Fund
under Project P16536-N04;  
part of this work was done while she visited the University of Toronto, 
which was made possible by Hector Levesque and Ray Reiter.

\end{document}